\documentclass[aps,prc,superscriptaddress,twoside,twocolumn,showpacs]{revtex4-1}
\usepackage{amsmath,amssymb}
\usepackage{mathtools}
\usepackage{bbold}
\usepackage[usenames, dvipsnames]{xcolor}
\usepackage{braket}
\usepackage{graphicx,bm}
\usepackage{slashed}
\usepackage{booktabs}
\usepackage{tensor}

\usepackage{multirow}

\usepackage{mwe}

\usepackage{changes}

\usepackage{enumerate}

\allowdisplaybreaks

\newcommand{\mce}{\mathcal{E}}
\newcommand{\dele}{\Delta \mathcal{E}}

\newcommand{\msun}{M_{\odot}}

\def\nicer{MR}
\def\ligo{M\Lambda}
\def\NS{{ \mathrm{NS} }}

\def\m{\mathcal}

\begin{document}

\title{Radius and equation of state constraints from massive neutron stars and GW190814}
\date{\today}

\author{Yeunhwan \surname{Lim} }
\email{ylim@mpi-hd.mpg.de}
\affiliation{Max-Planck-Institut f\"ur Kernphysik, Saupfercheckweg 1, 69117 Heidelberg, Germany}
\affiliation{Institut f\"ur Kernphysik, Technische Universit\"at Darmstadt, 64289 Darmstadt, Germany}
\affiliation{ExtreMe Matter Institute EMMI, GSI Helmholtzzentrum f\"ur Schwerionenforschung GmbH, 64291 Darmstadt, Germany}

\author{Anirban \surname{Bhattacharya} }
\email{anirbanb@stat.tamu.edu}
\affiliation{Department of Statistics, Texas A\&M University, College Station, TX 77843, USA}

\author{Jeremy W. \surname{Holt} }
\email{holt@physics.tamu.edu}
\affiliation{Cyclotron Institute, Texas A\&M University, College Station, TX 77843, USA}
\affiliation{Department of Physics and Astronomy, Texas A\&M University, College Station, TX 77843, USA}

\author{Debdeep \surname{Pati} }
\email{debdeep@stat.tamu.edu}
\affiliation{Department of Statistics, Texas A\&M University, College Station, TX 77843, USA}


\begin{abstract}
Motivated by the unknown nature of the $2.50-2.67\,\msun$ compact object in the binary merger event GW190814, we study the maximum neutron star mass based on constraints from low-energy nuclear physics, neutron star tidal deformabilities from GW170817, and simultaneous mass-radius measurements of PSR J0030+045 from NICER. Our prior distribution is based on a combination of nuclear modeling valid in the vicinity of normal nuclear densities together with the assumption of a maximally stiff equation of state at high densities, a choice that enables us to probe the connection between observed heavy neutron stars and the transition density at which conventional nuclear physics models must break down. We demonstrate that a modification of the highly uncertain supra-saturation density equation of state beyond 2.64 times normal nuclear density is required in order for 
chiral effective field theory models to be consistent with current neutron star observations and the existence of $2.6\,\msun$ neutron stars. We also show that the existence of very massive neutron stars strongly impacts the radii of $\sim 2.0\,\msun$ neutron stars (but not necessarily the radii of $1.4\,\msun$ neutron stars), which further motivates future NICER radius measurements of PSR J1614-2230 and PSR J0740+6620.
\end{abstract}


\maketitle

{\it Introduction---}
Recently, the LIGO/Virgo Collaboration (LVC) has reported measurements \cite{abbott20a} of gravitational waves resulting from a $2.50-2.67\,\msun$ ``mass-gap'' object \cite{Ozel10} in binary coalescence with a heavy ($22.2-24.3)$\,$\msun$ companion black hole. Not only are the mass ratio of $q=0.112^{+0.008}_{-0.009}$ and inferred merger rate of $1-23\,{\rm Gpc}^{-3}{\rm yr}^{-1}$ challenging to describe \cite{abbott20a,Zevin:2020gma,Vattis:2020iuz} with traditional binary evolution models, but taken at face value, the mass-gap secondary object in the observation represents the discovery of either the heaviest known neutron star (NS) or the lightest known black hole (BH). Neither the absence of a measurable tidal deformation signature in the gravitational waveform nor the absence of an electromagnetic counterpart would be unexpected \cite{foucart13,abbott20a} for a NSBH merger at the extreme mass ratio reported in GW190814. However, equation of state inferences \cite{abbott18} based on the binary neutron star merger event GW170817 and properties of its electromagnetic counterpart  \cite{bauswein17,margalit17,ruiz18,rezzolla18,radice18} suggest that the maximum mass of slowly rotating neutron stars, $M_{\rm max}^{\rm TOV}$, is bounded above by $M_{\rm max}^{\rm TOV} \lesssim 2.3\,\msun$. Population studies \cite{ozel12,alsing18,farr20} of known galactic pulsars exhibit a double-peaked mass distribution with extended high-mass tail, leading to the less restrictive bound \cite{abbott20a} $M_{\rm max}^{\rm TOV} = 2.25^{+0.81}_{-0.26}\,\msun$ with associated probability of 29\% that the GW190814 secondary mass lies below $M_{\rm max}^{\rm TOV}$. On the other hand, rapid uniform rotation \cite{cook94,sedrakian20,most20} allows for stable neutron star mass configurations about 20\% above the maximum nonrotating mass, though dissipation and electromagnetic spindown before merger would likely reduce the GW190814 secondary's spin, which was essentially unconstrained by observation \cite{abbott20a}.

Equation of state models used to interpret the nature of the GW190814 secondary \cite{abbott20a,sedrakian20,tews21}, or more generally to predict maximum neutron star masses, incorporate a variety of high-density extrapolations, including piecewise polytropes \cite{hebeler10prl,raithel16}, speed of sound parameterizations \cite{tews18,capano20,tan20}, spectral representations \cite{lindblom10,fasano19}, parameter-free inference via Gaussian processes \cite{landry18,essick19,landry20}, or smooth extrapolations from low-density regions constrained by nuclear physics \cite{lim19a,fattoyev20}. In this work, our aim is to understand model-independent implications for the high-density nuclear equation of state under the assumption that the GW190814 secondary is a neutron star. For this purpose, we choose an equation of state prior distribution that at low-density is informed by nuclear theory and experiment and at high densities is maximally stiff. On the one hand this choice explores a more limited range of high-density extrapolations than previous studies, but on the other hand it enables robust predictions for the critical density at which conventional nuclear physics models break down. The transition density $n_t$ to the maximally-stiff equation of state is varied within the range $2-4n_0$, where $n_0 = 0.16$\,fm$^{-3} = 2.4 \times 10^{14}$\,g/cm$^3$ is nuclear saturation density. We then include likelihood functions that incorporate recent radius and tidal deformability measurements of $\sim1.4\,\msun$ neutron stars as well as assumptions on the nature of the GW190814 secondary. From the posterior probability distribution, we explore the minimum transition density required to support slowly rotating $2.5-2.7\,\msun$ neutron stars and find that it lies in the range $n_t \sim (2.39-2.95) n_0$, which is below the central density of neutron stars with masses $M\sim 1.4\,\msun$ \cite{lim19a}. Nevertheless, we find that the existence of massive $\sim 2.6\,\msun$ neutron stars need not strongly constrain the bulk properties of typical lighter neutron stars, but instead only the softest equations of state with small radii and small tidal deformabilities are excluded. In contrast, we show that the radii of heavy neutron stars ($M \sim 2.0\,\msun$) are positively correlated with the maximum neutron star mass and may offer unique insights into the nature of ultra-dense matter \cite{han20}.

{\it Bayesian modeling of the neutron star equation of state---}
Experimentally measured nuclear binding energies and bulk oscillation modes constrain \cite{dutra12,dutra14} the nuclear equation of state around normal nuclear density $n_0$ for matter consisting of nearly equal numbers of neutrons and protons. Neutron-rich matter, on the other hand, is challenging to produce and study in the laboratory, and therefore the principal nuclear physics constraints on the neutron star equation of state rely in one way or another on nuclear theory models, which nowadays have a firm foundation in chiral effective field theory  \cite{weinberg79,epelbaum09rmp,machleidt11,drischler21h}, the low-energy realization of quantum chromodynamics. Previously, we have incorporated constraints from chiral effective field theory \cite{holt17prc,holt18} and nuclear experiments \cite{PhysRevC.85.035201,Lattimer:2012xj} to construct \cite{lim18a,lim19a} Bayesian posterior probability distributions for the neutron star equation of state parameterized in terms of a Taylor series expansion in the Fermi momentum $k_F \sim n^{1/3}$:
\begin{equation}
\label{eq:fun}
\begin{aligned}
\mathcal{E}(n,\delta) =  \frac{1}{2m}(\tau_n + \tau_p)
  + [1-\delta^2] f_s(n) + \delta^2 f_n(n) \,,
\end{aligned}
\end{equation}
where $\tau_n$ $(\tau_p)$ is the neutron (proton) kinetic energy density, $\delta = (n_n-n_p)/(n_n+n_p)$ is the proton-neutron asymmetry parameter, and the potential energy density is expanded as 
\begin{equation}
\label{eq:fns}
f_s(n) = \sum_{i=0}^3 a_i\, n^{(2+i/3)} \,, 
\quad
f_n(n) = \sum_{i=0}^3 b_i\,n^{(2+i/3)}\,.
\end{equation}
In all of our neutron star structure models, we construct a realistic outer and inner crust using the same parameters $(\vec a, \vec b)$ in a unified way implementing the liquid drop model as explained in more detail in Ref.\ \cite{lim17}. When these models were extrapolated to high densities, the maximum neutron star mass was found to be $M_{\rm max}^{\rm TOV} \simeq 2.3\,\msun$. Numerous other works have employed chiral effective field theory to study the neutron matter equation of state \cite{hebeler2010,coraggio13,gezerlis13,carbone14,drischler16,tews16,Piarulli:2019pfq,drischler2019,drischler20}, neutron star radii \cite{hebeler10prl,raaijmakers19,capano20}, tidal deformabilities \cite{annala18,most18,tews18gw}, and moments of inertia \cite{Lim2019b,greif20}. In describing the properties of the heaviest neutron stars, whose central densities can reach up to $n=5-10n_0$, all of these models perform extrapolations into regions where the composition and dynamics are poorly understood. 

To explore the widest range of maximum neutron star masses, we extend this previous model for the equation of state probability distribution to include a transition to the maximally-stiff equation of state consistent with relativity, defined when the speed of sound is equal to the speed of light. The transition density $n_t$ is taken to have a uniform prior in the range $2n_0 < n_t < 4n_0$. A transition density beyond $4n_0$ produced only minor differences in the equation of state prior. Formally, we employ a second-order phase transition where the phase transition starts at $\mce=\mce_1$ and ends at $\mce=\mce_2$, where $\dele = \frac{\mce_1}{10}$.

The approach described above defines the prior distribution $\pi(\cdot)$, and we construct Bayesian posterior probability distributions as follows. Having observed neutron star tidal deformabilities associated with GW170817 \cite{abbott17,abbott18,de2018,tidal170817} and simultaneous mass-radius measurements  \cite{Miller2019,Riley2019} of PSR J0030+045 from the NICER mission, the posterior distribution of $\theta$ is proportional to $\m L^{R\Lambda}(\theta) \, \pi(\theta)$, where $\m L^{R\Lambda}(\theta) = \m L^{\nicer}(\theta) \m L^{\ligo}(\theta)$ is the likelihood function of $\theta$. Here $\m L^{\nicer}(\theta)$ denotes the likelihood contribution from two NICER mass-radius measurements, which we combine with statistically equal weights, and $\m L^{\ligo}(\theta)$ denotes the joint $\{M_1, \Lambda_1; M_2, \Lambda_2\}$ posterior distribution from the LIGO analysis of GW170817. Since these are two independent measurements, the combined likelihood assumes a product form. 

We now detail the construction of the $\m L^{\nicer}$ and the $\m L^{\ligo}$ likelihoods. Let $R_\theta(M)$ denote the (unique) radius-versus-mass curve corresponding to the set of parameters in $\theta$. Each $R_\theta(\cdot)$ curve has its own maximum mass $M_\theta^{\max}$ above which the neutron star would undergo gravitational collapse, and hence the domain of $R_\theta(\cdot)$ is $(M_\theta^{\min}, M_\theta^{\max})$, where in all cases we take $M_\theta^{\min} = 1.0\,\msun$.  Although the correlated uncertainty corresponding to either NICER measurement resembles a tilted ellipse, a closer inspection of the contour plots reveal departures from normality. As a result, we build a non-parametric likelihood using a bivariate kernel density estimator (kde). To that end, we denote by $\{(M_i, R_i), i=1, \ldots, n_1\}$ and $\{(M_i, R_i), i=n_1+1, \ldots, n_2\}$ the posterior samples $(M,R)$ obtained from Fig.\ 7b of Ref.\ \cite{Miller2019} and Fig.\ 19 (``ST+PST'') of Ref.\ \cite{Riley2019} respectively. We fit a bivariate kernel density estimator $\widehat{f}$ to the pooled samples $\{(M_i, R_i), i=1, \ldots, n_2\}$. We used the \texttt{R} package \texttt{ks} to fit the kde, employing a  bivariate Gaussian kernel and the smoothed cross-validation estimator for the bandwidth matrix. Then, we consider an ``average" of this fitted density over an $R(M)$ curve as the corresponding likelihood, i.e.,
\begin{equation}\label{eq:int_lik}
\m L^{\nicer}(\theta) = \int_{M_\theta^{\min}}^{M_\theta^{\max}} \widehat{f} \big(M, R_\theta(M) \big) \,\frac{dM}{M_\theta^{\max}-M_\theta^{\min}}.  \end{equation}
The construction of the $\m L^{\ligo}$ likelihood is similar, except that we fit a quadvariate kde to the posterior samples of $\{M_1, \Lambda_1; M_2, \Lambda_2\}$ with a quadvariate Gaussian with a smoothed cross-validation estimator for the bandwidth matrix.

\begin{figure}
\includegraphics[scale=0.28]{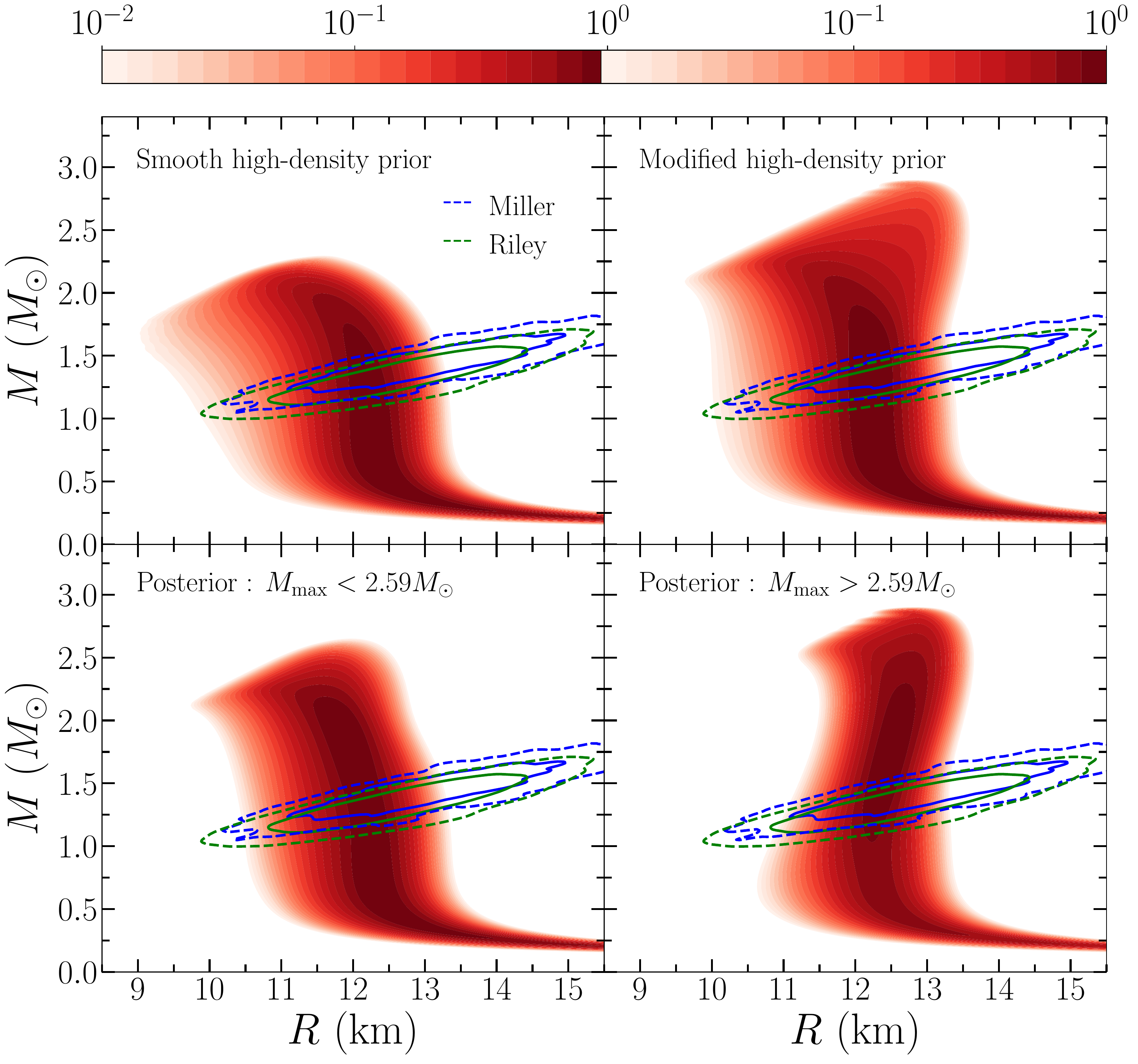}
\caption{(Color online) Mass and radius probability distributions for the (top-left) prior with smooth high-density extrapolation, (top-right) prior with maximally-stiff high-density extrapolation, (bottom-left) posterior not supporting $\sim 2.6\,\msun$ neutron stars, and (bottom-right) posterior supporting $\sim2 .6\,\msun$ neutron stars. The green \cite{Riley2019} and blue \cite{Miller2019} contours represent the NICER 68\% (solid) and 95\% (dashed) credibility bands.}
\label{fig:mrdist}
\end{figure}

We incorporate the secondary ``mass-gap'' object into our likelihood function using the GW190814 posterior mass samples \cite{gw190814pos}. Although the distribution of the secondary mass $M_\mathrm{s}$ in Fig.\ 4 of Ref.\ \cite{abbott20a} resembles a Gaussian, a Kruskal-Wallis test rejected the hypothesis of Gaussianity. Hence, we used a nonparametric density estimate $\widehat{f}_{M_\mathrm{s}}$ using univariate kde to approximate the distribution of $M_\mathrm{s}$, again with a smoothed cross-validation estimator for the bandwidth. Hence, for a given value of $\theta$ from the equation of state, the secondary object is realizable as a neutron star with probability given by 
\begin{eqnarray}\label{eq:s_NS}
 \mathbb{P}(M_\mathrm{s} \leq M_\theta^{\max}) = \int_{M_\theta^{\min}}^{M_\theta^{\max}} \widehat{f}_{M_\mathrm{s}}(M) dM.
\end{eqnarray}
Eq.\,\eqref{eq:s_NS} then defines the likelihood of the object assuming it to be a neutron star, denoted $\mathcal{L}_\mathrm{s}^{\NS}(\theta)$, which when multiplied with $\m L^{R\Lambda}(\theta)$ gives the overall likelihood for $\theta$. 
Similarly, if we assume the object is a black hole, then the likelihood involves the probability given by $\mathcal{L}_\mathrm{s}^{\mathrm{BH}}(\theta) :\,= 1-   \mathbb{P}(M_\mathrm{s} \leq M_\theta^{\max})$. While alternative exotic compact objects have been theorized to exist (see Ref.\ \cite{cardoso19} for a recent review), here we consider the GW190814 secondary to be either a black hole or neutron star.

\begin{figure}
\includegraphics[scale=0.5]{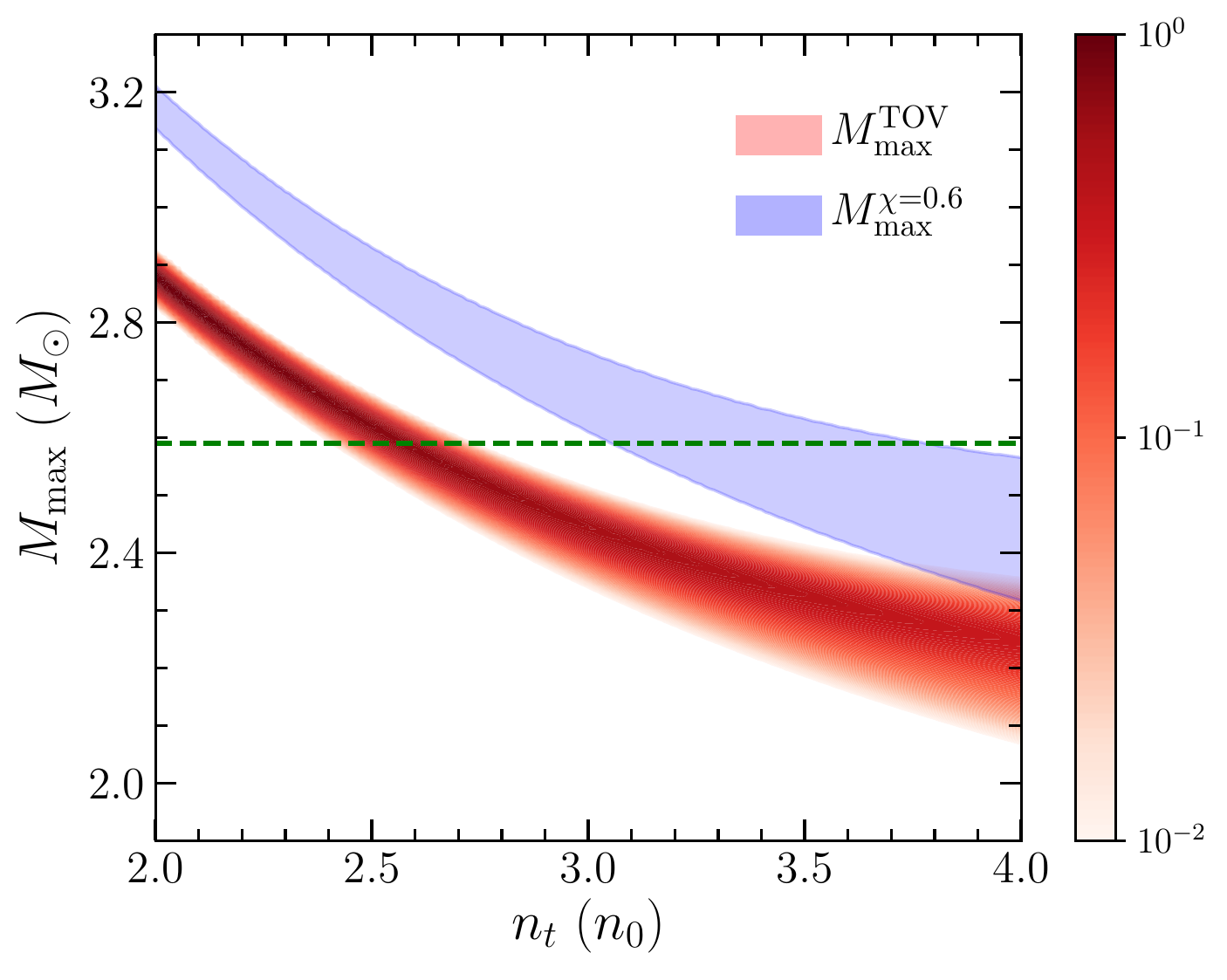}
\caption{(Color online) Probability distribution for the maximum neutron star mass (red) $M_{\rm max}^{\rm TOV}$ and (blue) $M_{\rm max}^{\chi}$ for $\chi = 0.6$ as a function of the transition density $n_t$ to the maximally-stiff high-density equation of state. The green dashed line lies at the central value, $2.59\,\msun$, of the GW190814 secondary.}
\label{fig:nt}
\end{figure}

{\it Results---}
In Fig.\ \ref{fig:mrdist} we show the mass and radius probability distributions based on the Bayesian analysis described above. In all subpanels of Fig.\ \ref{fig:mrdist}, the green and blue contours represent the $68\%$ (solid lines) and 95\% (dashed lines) credibility bands obtained from our kernel density estimators associated with the Riley {\it et al.}\,\cite{Riley2019} and Miller {\it et al.}\,\cite{Miller2019} analyses of NICER x-ray waveform data from PSR J0030+045. The top-left figure is our previous prior \cite{lim18a} with a smooth high-density extrapolation, and the top-right panel is our new prior with uniformly varying transition density $2n_0 < n_t < 4n_0$ to the maximally-stiff equation of state. We see that the modified high-density prior naturally leads to much larger maximum neutron star masses, up to $M_{\max}^{\rm TOV} = 2.9\,\msun$ for the lowest value of the transition density considered $n_t=2n_0$. We note that this new maximum neutron star mass of $M_{\max}^{\rm TOV} = 2.9\,\msun$ is almost certainly unphysical since it lies above the total mass $M_{\mathrm{tot}} \simeq 2.7\,\msun$ of the GW170817 remnant, which is expected \cite{2017Natur.551...80K} to have collapsed to a black hole after being supported initially through differential rotation.

The bottom-left and bottom-right panels of Fig.\ \ref{fig:mrdist} represent the posterior mass-radius probability distributions under the assumption that the secondary in GW190814 was a slowly rotating black hole or neutron star, respectively. Interestingly, we see that for typical neutron stars with masses $M\sim 1.4\,\msun$, the distribution of radii is not strongly different under the two interpretations of the GW190814 secondary. This is due to the fact that the bulk  properties of the average neutron star are strongly correlated \cite{0004-637X-550-1-426,lim18a,Tsang:2019mlz} with the pressure of beta-equilibrium matter at the density $n=2n_0$, which is close to the regime where nuclear physics places strong constraints on the equation of state. However, we do observe that the existence of massive $(2.5-2.6)\,\msun$ neutron stars would rule out the softest equations of state. In particular, neglecting $\m L^{R\Lambda}(\theta)$, our previous finding \cite{lim18a} for the radius of a $1.4\,\msun$ neutron star at the $95\%$ credibility level was $10.3\,{\rm km} \le R_{1.4} \le 12.9$\,km. Including the new kde constraints from the two NICER and GW170817 analyses now give at the $95\%$ credibility level $10.9\,{\rm km} \le R_{1.4} \le 12.9$\,km under the assumption $\mathcal{L}_\mathrm{s}^{\mathrm{BH}}(\theta)$ and $11.5\,{\rm km} \le R_{1.4} \le 13.0$\,km under the assumption $\mathcal{L}_\mathrm{s}^{\mathrm{NS}}(\theta)$.

\begin{figure}[t]
\includegraphics[scale=0.56]{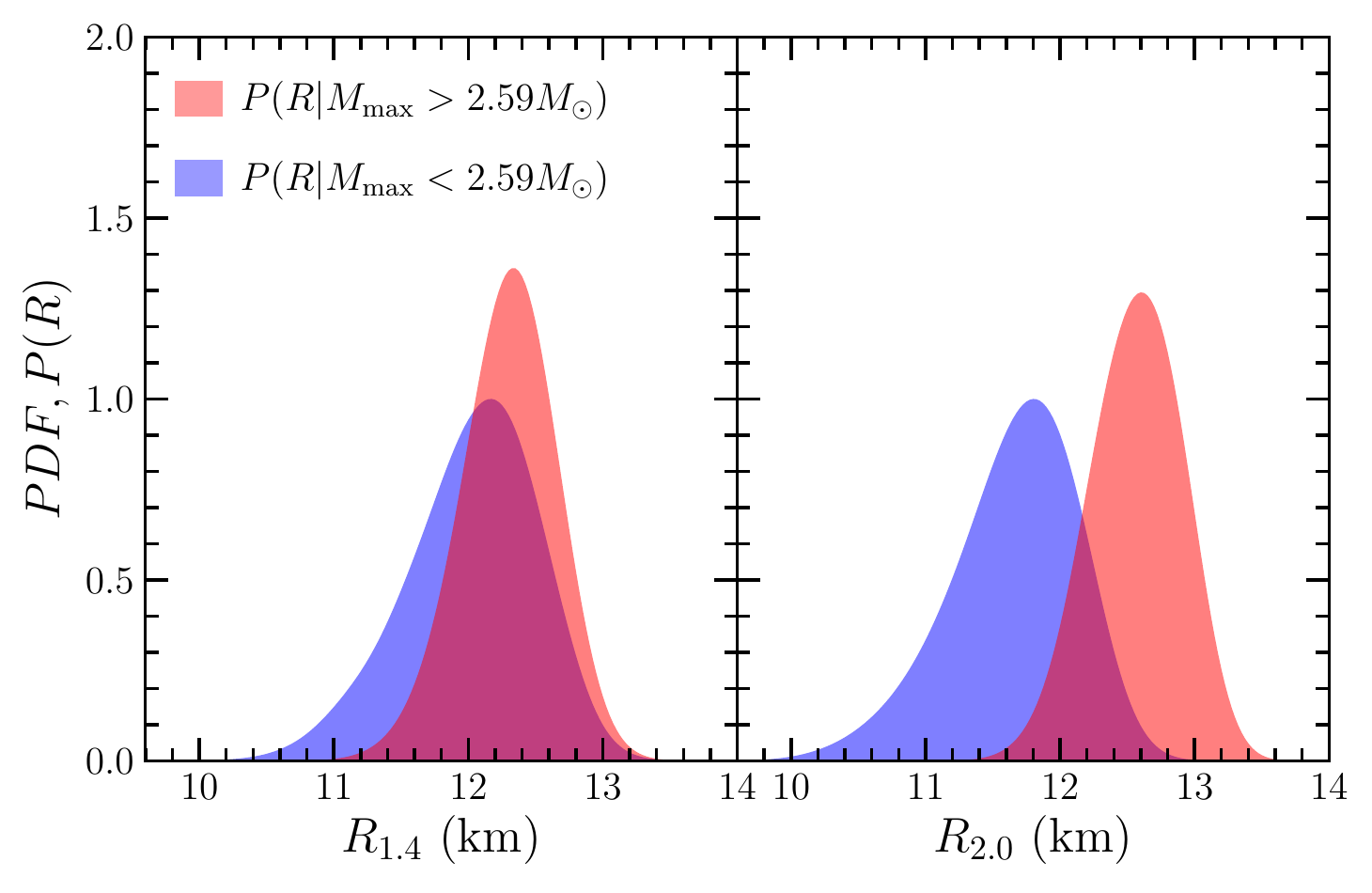}
\caption{(Color online) Radius distributions for a $1.4\,\msun$ neutron star (left) and $2.0\,\msun$ neutron star (right) under the two assumptions that the GW190814 secondary was a black hole (blue) or a neutron star (red).}
\label{fig:m214rdist}
\end{figure}

In Fig.\ \ref{fig:nt} we plot the maximum neutron star mass as a function of the transition density $n_t$ from the posterior distribution $\m L^{R\Lambda}(\theta) \, \pi(\theta)$, that is, without any assumptions regarding the nature of the GW190814 secondary. At the largest value of $n_t = 4n_0$, the maximum neutron star mass allowed in the present modeling is $2.32\,\msun$. In order to support $\{ 2.5, 2.6, 2.7\} \,\msun$ neutron stars, the transition density must be less than $\{2.95, 2.64, 2.39\} n_0$, respectively. Therefore, the relatively soft equations of state predicted by chiral effective field theory would have to become fairly stiff soon after their natural breakdown scale in the range $1-2n_0$ (see also Refs.\ \cite{essick20,drischler20r} for recent related discussions). We note that since our high-density equation of state is maximally stiff, the above constraints must be satisfied for any other choice of high-density extrapolation. In addition, in Fig.\ \ref{fig:nt} we plot the relation between $M^\chi_{\rm max}$ and $n_t$ under the assumption of rapid rotation $\chi=0.6$, which has recently been suggested \cite{most20} as the central value of the GW190814 secondary's spin under the assumption it was once a neutron star. For this calculation we employ the quasi-universal relation between $M_{\rm max}^{\rm TOV}$ and $M_{\rm max}^{\chi}$ for a uniformly-rotating neutron star with spin $\chi$ derived in Ref.\ \cite{breu16}. Although rapid rotation ($\chi = 0.6$) alone would be sufficient to stabilize $2.6\,\msun$ neutron stars for nearly all of our posterior samples, the unknown evolutionary path of GW190814 and its observationally unconstrained spin make it difficult to draw additional inferences at this time.

\begin{figure}
\includegraphics[scale=0.44]{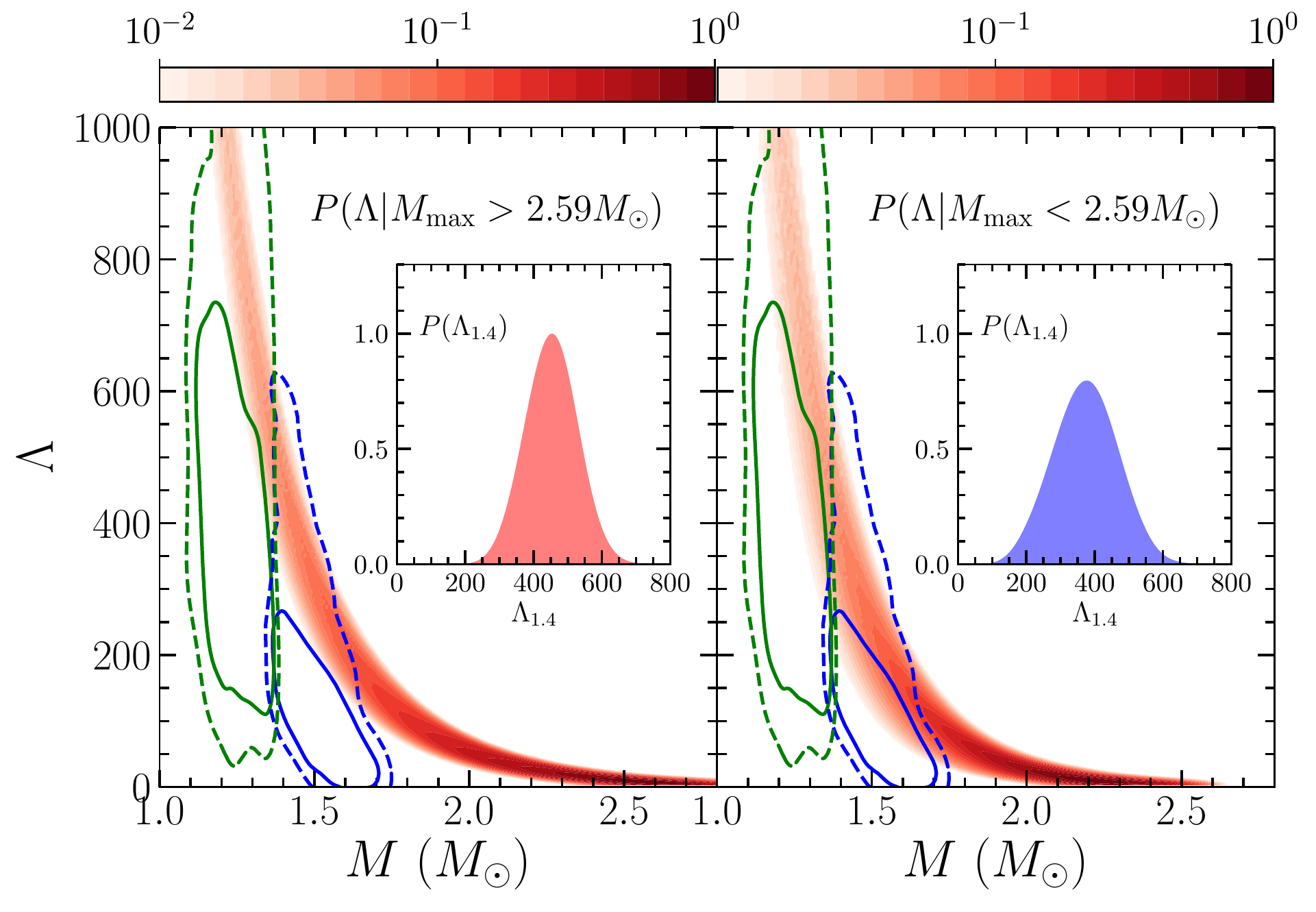}
\caption{(Color online) Probability distributions for the tidal deformability versus mass under the two assumptions $\mathcal{L}_\mathrm{s}^{\mathrm{NS}}(\theta)$ (left) and $\mathcal{L}_\mathrm{s}^{\mathrm{BH}}(\theta)$ (right). The blue and green contours show the 68\% (solid) and 95\% (dashed) marginal likelihoods associated with the primary and secondary in GW170817, respectively. Inset: probability distributions for $\Lambda_{1.4}$ in the two competing scenarios.}
\label{fig:lmdist}
\end{figure}

We see from Fig.\ \ref{fig:mrdist} that in contrast to typical neutron stars with $M=1.4\,\msun$, heavy neutron stars have significantly different radius probability distributions under our two assumptions for the GW190814 likelihood, $\mathcal{L}_\mathrm{s}^{\mathrm{BH}}(\theta)$ and $\mathcal{L}_\mathrm{s}^{\mathrm{NS}}(\theta)$. In Fig.\ \ref{fig:m214rdist} we show the posterior probability distributions for the radius of a $1.4\,\msun$ neutron star (left) and $2.0\,\msun$ neutron star (right) under the two assumptions that the GW190814 secondary was a black hole (blue) or a neutron star (red). The notation $P(R | M_{\max} > 2.59\,\msun)$ corresponds to the likelihood assumption $\mathcal{L}_\mathrm{s}^{\mathrm{NS}}(\theta)$ and likewise $P(R | M_{\max} < 2.59\,\msun)$ corresponds to $\mathcal{L}_\mathrm{s}^{\mathrm{BH}}(\theta)$.
The stiff equations of state needed to support the heaviest neutron stars produce a radius distribution for $M=2.0\,\msun$ that is narrower and peaked at larger central values compared to softer equations of state. In contrast, for typical $1.4\,\msun$ neutron stars the two radius distributions under the $\mathcal{L}_\mathrm{s}^{\mathrm{NS}}(\theta)$ and $\mathcal{L}_\mathrm{s}^{\mathrm{BH}}(\theta)$ posteriors are very strongly overlapping.
We find at the 95\% credibility level that the radius of a $2.0\,\msun$ neutron star is $10.5\,{\rm km} < R_{2.0} < 12.5$\,km under the assumption $\mathcal{L}_\mathrm{s}^{\mathrm{BH}}(\theta)$ and $11.8\,{\rm km} < R_{2.0} < 13.2$\,km under the assumption $\mathcal{L}_\mathrm{s}^{\mathrm{NS}}(\theta)$.

In Fig.\ \ref{fig:lmdist} we show the two posterior probability distributions for the tidal deformability as a function of mass under the two assumption for the likelihood $\mathcal{L}_\mathrm{s}^{\mathrm{BH}}(\theta)$ and $\mathcal{L}_\mathrm{s}^{\mathrm{NS}}(\theta)$. In both subpanels the blue and green contours denote the 68\% (solid lines) and 95\% (dashed lines) marginal likelihoods (e.g., $\int \mathcal{L}(M_1, \Lambda_1; M_2, \Lambda_2)\, d M_2 \, d \Lambda_2$ for the blue contours) from our quadvariate kde associated with the primary and secondary components, respectively, of GW170817. As insets to Fig.\ \ref{fig:lmdist}, we show the tidal deformability of a typical $1.4\,\msun$ neutron star under the assumptions that the GW190814 secondary was a neutron star (red) or black hole (blue). Our previous 95\% credibility interval in Ref.\ \cite{lim18a} was found to be $140 < \Lambda_{1.4} < 520$. From the new posterior distribution including NICER and GW170817 measurements as well as the assumption $\mathcal{L}_\mathrm{s}^{\mathrm{BH}}(\theta)$, we find $190 < \Lambda_{1.4} < 550$. Under the opposite scenario, $\mathcal{L}_\mathrm{s}^{\mathrm{NS}}(\theta)$, we likewise find $310 < \Lambda_{1.4} < 590$ at the 95\% credibility level.

{\it Summary---}
The existence of heavy neutron stars with masses $2.5-2.6\,\msun$ are a challenge to explain with equations of state smoothly extrapolated from the low-density regime ($1-2n_0$) constrained by nuclear physics to the highest-density regime ($5-10n_0$) encountered in neutron star cores. We have demonstrated that a modification of the highly uncertain supra-saturation density equation of state above a transition density of $n_t=2.64\,n_0$ is necessary for the support of slowly rotating $\sim 2.6\,\msun$ neutron stars while maintaining consistency with state-of-the-art nuclear theory modeling within the framework of chiral effective field theory, nuclear experiments involving medium-mass and heavy isotopes, as well as current observations of neutron star radii and tidal deformabilities. In our modeling, the existence of very heavy neutron stars ($2.5-2.6\,\msun$) is not strongly correlated with the radii of typical $1.4\,\msun$ neutron stars but is correlated with the radii of $\sim 2.0\,\msun$ neutron stars. We suggest that NICER measurements of e.g., the PSR J1614-2230 or PSR J0740+6620 radius may therefore provide useful and strong constraints on the nuclear equation of state at supra-saturation density.

\acknowledgments
Y. Lim was supported by the Max Planck Society and the Deutsche Forschungsgemeinschaft (DFG, German Research Foundation) -- Project ID 279384907 -- SFB 1245. Dr.\ Pati and Dr.\ Bhattacharya acknowledge support from NSF DMS (1854731, 1916371) and NSF CCF 1934904 (HDR-TRIPODS). In addition, Dr.\ Bhattacharya acknowledges NSF CAREER 1653404 for supporting this project. The work of J.\ W.\ Holt is supported by the National Science Foundation under Grant No.\ PHY1652199 and by the U.\ S.\ Department of Energy National Nuclear Security Administration under Grant No.\ DE-NA0003841. Portions of this research were conducted with the advanced computing resources provided by Texas A\&M High Performance Research Computing.

\bibliographystyle{apsrev4-1}


%

\end{document}